\documentclass[12pt, pre, aps, showpacs, preprint]{revtex4}

\pdfoutput=1
\usepackage{amsmath}
\usepackage{amsfonts}
\usepackage[pdftex]{graphicx}
\usepackage{mathrsfs}

\begin{document}

\title{Dense Sphere Packings from Optimized Correlation Functions}
\author{Adam B. Hopkins}
\affiliation{Department of Chemistry; Princeton University, Princeton, New 
Jersey 08544}

\author{Salvatore Torquato}
\affiliation{Department of Chemistry, Princeton Institute for the Science and 
Technology of Materials, Program in Applied and Computational Mathematics, 
Princeton Center for Theoretical Science; Princeton University, Princeton, 
New Jersey 08544; School of Natural Sciences, Institute for Advanced Study; 
Princeton, New Jersey 08544}

\author{Frank H. Stillinger}
\affiliation{Department of Chemistry; Princeton University, Princeton, New 
Jersey 08544}

\begin{abstract}
Elementary smooth functions (beyond contact) are employed to construct pair 
correlation functions that mimic jammed disordered sphere packings. Using the 
$g_2$-invariant optimization method of Torquato and Stillinger 
[J. Phys. Chem. B {\bf 106}, 8354, 2002], parameters in these functions are 
optimized under necessary realizability conditions to maximize the packing 
fraction $\phi$ and average number of contacts per sphere $Z$. A pair 
correlation function that incorporates the salient features of a disordered 
packing and that is smooth beyond contact is shown to permit a $\phi$ of 
0.6850: this value represents a 45\% reduction in the difference between
the maximum for congruent hard spheres in three dimensions, $\pi/\sqrt{18}
\approx$ 0.7405, and 0.64, the approximate fraction associated with maximally 
random jammed (MRJ) packings in three dimensions. We show that, surprisingly, 
the continued addition of elementary functions consisting of smooth sinusoids 
decaying as $r^{-4}$ permits packing fractions approaching $\pi/\sqrt{18}$. A 
translational order metric is used to discriminate between degrees of order 
in the packings presented. We find that to achieve higher packing fractions, 
the degree of order must increase, which is consistent with the
results of a previous study 
[Torquato et al., Phys. Rev. Lett. {\bf 84}, 2064, 2000].  
\end{abstract}

\maketitle

\section{Introduction and background}
Packing problems address the various arrangements of a set (finite or 
infinite) of non-overlapping objects in a space of given dimension \cite{CSSPLG1999,Hales2005a, TorquatoRHM2002}.
Often, one seeks to arrange the objects in such a way as to 
optimize a certain statistical or bulk property, e.g. the number density 
$\rho$ of objects (or equivalently the packing fraction $\phi$, the fraction 
of space covered by the objects' interiors). This paper is concerned with 
optimizing the packing fraction $\phi$ and average number of contacts per sphere 
(average kissing number) $Z$ for packings of congruent spheres in three-dimensional 
Euclidean space while maintaining an assumed functional form 
for the pair correlation function (defined below) of a ``random" packing. 

Random packings of three-dimensional 
hard spheres have been studied by scientists to better understand everything 
from heterogeneous materials to liquids to granular media (like sand) to living cells
\cite{TorquatoRHM2002,To98,Edwards,Chaikin,Zohdi}. In parallel to 
the concept of a maximum packing fraction for periodic (crystalline) packings, it had 
been assumed that a (different) maximum packing fraction could be defined for random 
packings, referred to as ``random close packing'' (RCP) 
\cite{Bernal1959a,Scott1960a}. However, while there is a proved maximum 
packing fraction for hard-sphere periodic packings achieved via an FCC lattice or one 
of its stacking variants \cite{Hales2005a}, the RCP state has been shown to 
be ill-defined \cite{TTD2000a,TS2008a}.

Experimental packings of oiled steel ball bearings originally led to the idea 
that mechanically stable random packings of identical 
spheres could not exhibit packing fractions exceeding 0.64 or declining past 
0.60 \cite{Scott1960a,SK1969a}. Mathematically constructed models 
\cite{Finney1970a} and early computer simulations \cite{Finney1970b} seemed 
to support these conclusions, though later work demonstrated that the 
limiting packing fractions obtained were highly dependent on the packing 
methods \cite{TTD2000a,TorquatoRHM2002}. These methods included, for example, 
lightly vibrating a container filled with spheres in either horizontal or 
vertical motions, rolling spheres one by one into a container \cite{PNW1997a}, and simulating 
the compression of a hard-sphere gas \cite{LS1990a,LSP1991a}.

Changing the method of packing via molecular dynamics simulations showed that 
densities past 0.64 are realizable \cite{TTD2000a,TTD2000b,Ka02}. 
However, it has yet to be demonstrated from a theoretical basis, i.e., 
without resort to experiment or computer simulation, that indeed there is no 
maximum density limit for ``random" packings apart from the proved limit for 
periodic packings. This paper provides such a theoretical basis by extending 
previous optimization studies of $g_2$-invariant processes (defined below) 
\cite{TS2002a,TS2006a} to a broader class of disordered packings.

Following previous work by two of us, a statistically homogeneous and 
isotropic packing is defined to be disordered if its pair correlation 
function $g_2(r)$ in $d$-dimensional Euclidean space $\mathbb{R}^d$ decays to 
unity faster than $r^{-d-\epsilon}$ for some $\epsilon > 0$ \cite{TS2006a}. 
Each packing corresponds to a unique $g_2(r)$, a function 
proportional to the probability density of finding a separation $r$ between 
any two sphere centers and normalized such that it takes the value of unity 
when no spatial correlations are present. The precise definition of 
``disordered" via the pair correlation function takes the place of the 
imprecise term ``random" hereafter.

The essential ideas behind our approach were actually laid out in our earlier 
work \cite{TS2002a,TS2006a}. In Ref. \cite{TS2002a}, the main objective was 
to study disordered packings in which short-range order was controlled using 
so-called $g_2$-invariant processes. A \textit{$g_2$-invariant process} is 
one in which a given pair correlation function $g_2(r)$ remains invariant for 
all $r$ as packing fraction varies over the range of densities
\begin{equation}
0 \leq \phi \leq  \phi_*.
\label{g2invariantRange}
\end{equation}
The terminal packing fraction $ \phi_*$ is the maximum achievable for the 
$g_2$-invariant process subject to satisfaction of the nonnegativity of 
$g_2(r)$ and the structure factor $S(k)$, i.e.,
\begin{equation}
g_2(r) \geq 0 \,\,\,\,\, \forall \,\,r \geq 0,
\label{condition1}
\end{equation}
\begin{equation}
S(k) \geq 0 \,\,\,\,\, \forall \,\,k \geq 0.
\label{condition2}
\end{equation}
For a statistically homogeneous and isotropic packing at number density 
$\rho$, $S(k)$ is related to the Fourier transform of the 
\textit{total correlation function} $h(r) \equiv g_2(r) - 1$, by
\begin{equation}
S(k) = 1 + \rho \tilde{h}(k),
\label{structFact1}
\end{equation}
where $\tilde{h}(k)$ represents the Fourier transform of $h(r)$. In three 
dimensions, this can be written as
\begin{equation}
S(k) = 1 + 4\pi\rho\int_0^{\infty}\frac{r\sin{kr}}{k}h(r)dr.
\label{structFact2}
\end{equation}
The optimization procedure described above was formulated for a hard-sphere 
packing, which requires the additional condition on $g_2$ that respects the 
nonoverlap constraint, i.e., 
\begin{equation}
g_2(r)=0  \qquad \mbox{for}\quad  0 \le r < D.
\label{condition}
\end{equation}  
When there exist sphere packings with a $g_2$ satisfying conditions 
(\ref{condition1}), (\ref{condition2}), and (\ref{condition}) for $\phi$ in 
the interval $[0, \phi_*]$, then a lower bound on the maximal 
packing fraction is given by $\phi_{max} \ge \phi_*$. It is noteworthy that this 
optimization problem for sphere packings is an infinite-dimensional linear 
program, which is the dual of the primal  linear program devised by Cohn and 
Elkies \cite{CE2003a} to obtain upper bounds on the maximal packing fraction
\cite{TS2006a}. We will comment further on this connection in the 
conclusions. Finally, we note here that the results of the 
optimization of the pair correlation function given in Ref. \cite{TS2002a} has found application 
in describing small-scale convective structural features of the solar surface \cite{Be04}.

The nonnegativity conditions (\ref{condition1}) and (\ref{condition2}) are 
necessary, but generally not sufficient, for a pair correlation function at a 
given density  to be realizable by  a point process \cite{KLS2007a}. A third 
condition, obtained by Yamada \cite{Yamada1961a} and not included in the optimization procedure described above,
constrains $\sigma^2(A) \equiv \langle (N(A)-\langle N(A)\rangle)^2\rangle$,
which is the variance in the number of points $N(A)$ contained within a window $A \in \mathbb{R}^d$; 
\begin{equation}
\sigma^2(A) \geq \theta(A)\big[1 - \theta(A)\big],
\label{condition3}
\end{equation}
where $\theta(A)$ is the fractional part of the expected number of points 
contained within the window. The number variance associated with a spherical
window of radius $R$ for a statistically homogeneous point process in $d$ dimensions  
can be written  as follows \cite{TS2003a}:
\begin{equation}
\sigma^2(R) = \rho v_1(R)\big[1 + 
\rho\int_{\mathbb{R}^d}h(\textbf{r})\alpha_2^{int}(r;R)d\textbf{r}\big] \geq \theta(R)\big[1 - \theta(R)\big],
\label{sigmaSquared}
\end{equation}
where $v_1(R)$ is the volume of the window and $\alpha_2^{int}(r;R)$ the 
intersection volume of two windows of radius $R$ (whose centers are separated by 
$\textbf{r}$) divided by $v_1(R)$. This additional condition (\ref{sigmaSquared}) on the pair correlation
function is always satisfied for statistically homogeneous and isotropic packings with sufficiently large 
windows in dimensions greater than 1 \cite{TS2006a}. In all cases that have 
been studied, this condition is satisfied for all $R$ if the first two conditions are 
satisfied \cite{TS2006a,SST2008a}.

While conditions (\ref{condition1}), (\ref{condition2}), and 
(\ref{condition3}) are necessary for the realizability of point processes, 
along with incorporation in $g_2(r)$ of the core exclusion feature, they 
appear to be rather strong conditions for realizability of sphere packings, 
especially as the space dimension increases \cite{TS2006a}. For example, a 
method to construct disordered packing configurations that realize test 
$g_2$'s meeting the conditions and incorporating the features of core 
exclusion and contact pairs [Eqs. (\ref{g1}) and (\ref{g2})] has been 
successful \cite{CTS2003a,UST2006a}
in two and three dimensions. No example in three dimensions or greater of an 
unrealizable $g_2$ satisfying the conditions and incorporating the core 
exclusion feature is currently known.

In Ref. \cite{TS2002a}, a five-parameter test family of $g_2$'s incorporating 
features of core exclusion, contact pairs, and damped oscillatory short-range 
order beyond contact [Eqs. (\ref{g1}), (\ref{g2}), and (\ref{g3})] had been 
considered. The problem of finding the terminal packing fraction $\phi_*$ was 
posed as an optimization problem: maximize $\phi$ over the set of parameters 
subject to the first two realizability conditions (the third condition due to 
Yamada was not relevant). In this work, we consider a broader family of 
smooth $g_2$ test functions corresponding to disordered packings 
\cite{endnote2} and satisfying all three aforementioned conditions.

To demonstrate the absence of a theoretical upper limit on disordered 
packings, we show that terms decaying as $r^{-4}$, representative of a 
feature prominent in the pair correlation functions of maximally random 
jammed (MRJ) packings \cite{DST2005a}, allow for increased packing fraction 
for pair correlation functions satisfying the three conditions and 
incorporating the aforementioned features. A simple 11-parameter form 
consisting of the initial five-parameter form plus two sinusoids decaying as 
$r^{-4}$ permits a packing fraction of 0.6850. Using a translational order 
metric, we show that the pair correlation function with the highest packing 
fraction also exhibits the highest degree of order, which is consistent with 
the conclusions of a previous work \cite{TTD2000a}. Additionally we show the 
surprising result that the continued addition of terms decaying as $r^{-4}$ 
allows for packing fractions up to $\pi/\sqrt{18}$, indicating that, if the 
packings are realizable, the progression of disordered packings up to the 
maximum $\phi$ is a continuum, dependent only on the form and parameters of 
the functions employed. A qualitative description of a realizable disordered 
packing with smooth $g_2(r)$ and $\phi$ approaching 
$\pi/\sqrt{18}$ is provided in Section IV A.

\section{Optimization of $g_2$-invariant processes}

We begin by revisiting the optimization problem first examined in Ref. 
\cite{TS2002a}. We employ a more comprehensive search using simulated 
annealing to optimize the five parameters of the family of $g_2$'s presented 
in Ref. \cite{TS2002a} and find a higher terminal packing fraction $\phi_* =$ 0.64268. 
The three functions that comprise the five parameter family, $g_I(r)$, 
$g_{II}(r)$ and $g_{III}(r)$, capture the most salient properties of a 
disordered packing, including that the average number of spheres in contact 
is $Z$ and that no sphere centers may approach closer than a distance of one 
sphere diameter.

A Heaviside step function represents the spheres' hard core exclusion,
\begin{equation}
g_I = \Theta(r-1),
\label{g1}
\end{equation}
where we set the diameter of the spheres to be unity. The Heaviside step 
function $\Theta(x)$ is defined piecewise as
\begin{equation}
\Theta(x) =
\begin{cases}
0, & x < 0 \\
1, & x \geq 0. \\
\end{cases}
\label{Heaviside}
\end{equation}
A Dirac delta function represents pair contacts,
\begin{equation}
g_{II} = \frac{Z}{4\pi\rho}\delta(r-1),
\label{g2}
\end{equation}
with $Z$ the average number of contacts per sphere (average kissing number). 
An exponentially decaying sinusoid provides short-range oscillatory motion 
about unity:
\begin{equation}
g_{III} = \frac{A_1}{r}\exp{(-B_1r)}\sin(C_1r + D_1)\Theta(r-1),
\label{g3}
\end{equation}
with parameters $A_1$, $B_1$, $C_1$, $D_1$. The total pair correlation 
function $g_2(r)$ is then
\begin{equation}
g_2(r) = g_I(r) + g_{II}(r) + g_{III}(r).
\label{fiveForm}
\end{equation}
Constraining $B_1 > 0$, $Z \geq 0$ ensures a physical configuration, while 
constraining $C_1 \geq 0$, $0 \leq D_1 < \pi$ eliminates function duplicates 
without additionally constraining the range of the functional form.

\subsection{Maximizing packing fraction}

Packing fraction is maximized using the $g_2$-invariant method in 20,000 
independent runs of over 10,000 iterations each. Initial parameters are 
confined to the following bounds, and selected randomly before each run with 
exponentially decreasing probability from zero:
\begin{align*}
-50 & < A_1 < 100\\
0 & < B_1 < 10\\
0 & \leq C_1 < 50\\
0 & \leq D_1 < \pi\\
0 & \leq \, Z \,\,< 13\\
\label{g3initialBounds}
\end{align*}
Parameters are allowed to range outside of bounds, but in no cases of the 
20,000 did this occur.

Using $G_I(k)$, $G_{II}(k)$, $G_{III}(k)$ to represent $1/\rho$ times the 
second term on the right hand side of relation (\ref{structFact2}) with 
$h(r) = g_I(r) + g_{II}(r) + g_{III}(r) - 1$, the structure factor for the 
functions becomes
\begin{equation}
S(k) = 1 + \frac{Z\sin(k)}{k} + \rho(G_I(k) + G_{III}(k)),
\label{FourierRadialStructFact}
\end{equation}
with $G_{II}(k) = Z\sin(k)/\rho k$. The exact analytical forms for $G_I$, 
$G_{II}$, and $G_{III}$ are included in Appendix B.

The method to maximize $\phi$ relies upon a simple principal. If $S(K) = 0$
at some point $K$ and $G_I(K) + G_{III}(K) < 0$, while $S(k) \geq 0$ for all 
other points $k$, then $\phi$ is at a global maximum for the given five 
parameters, i.e., $\phi =  \phi_*$. Hence to maximize $ \phi_*$, $S(k)$ is 
analytically calculated from the pair correlation function in accordance with 
Eq. (\ref{FourierRadialStructFact}), and for each random step along one of 
the five parameters, if possible $\phi$ is chosen such that the structure 
factor is in accordance with this principal. If obeying the structure factor 
condition is not mathematically possible for the parameter set, if the pair 
correlation function $g_2(r)$ is not greater than or equal to zero for all 
$r$, or if the maximum $\phi$ for the set does not meet the standard 
temperature-dependent simulated annealing condition for accepting a move, the 
random step is rejected and a new step chosen.

\begin{figure}[ht]
\centering
\includegraphics[width = 4.0in,viewport = 25 30 720 
580,clip]{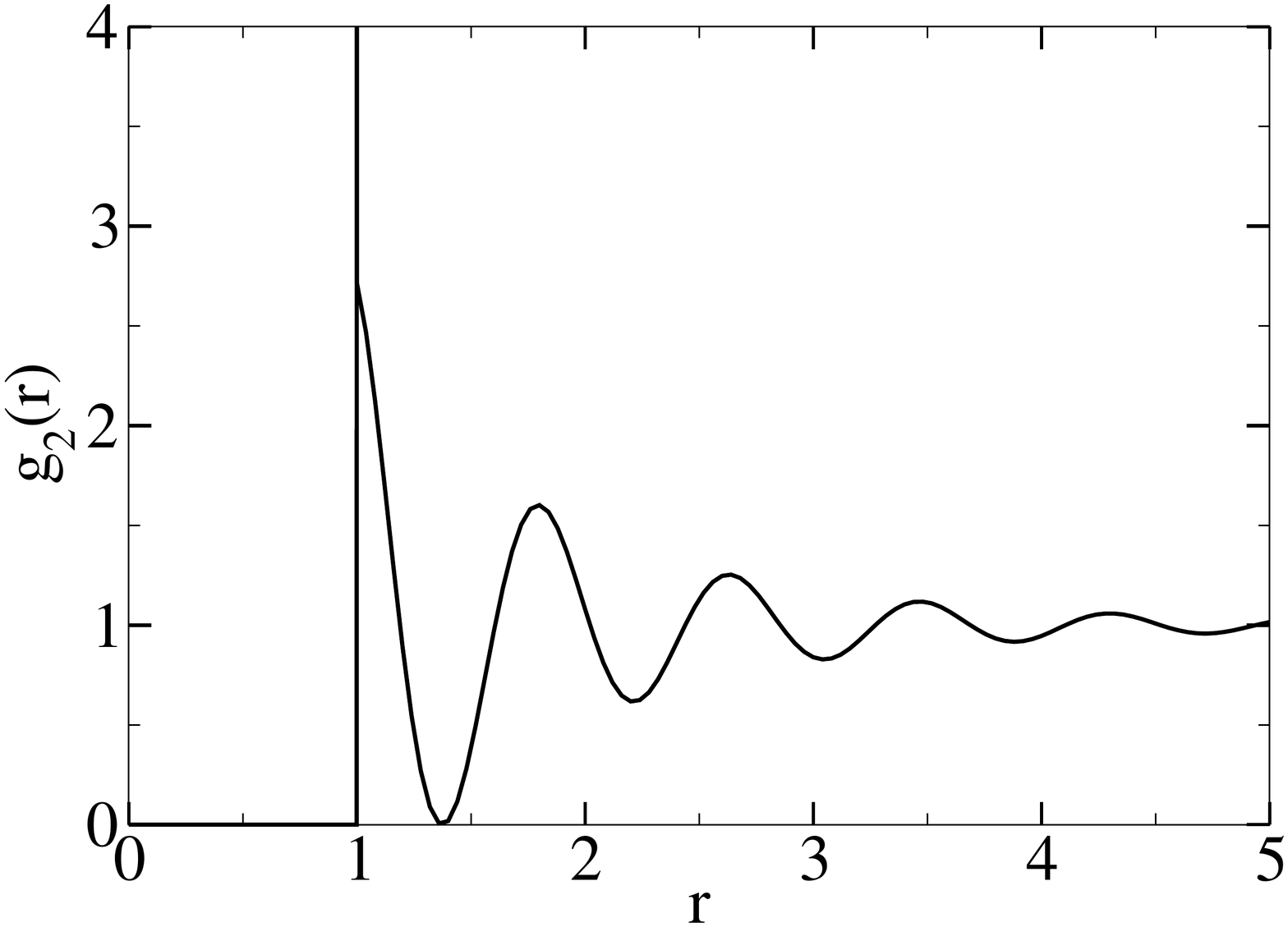}
\includegraphics[width = 4.0in,viewport = 25 30 720 
550,clip]{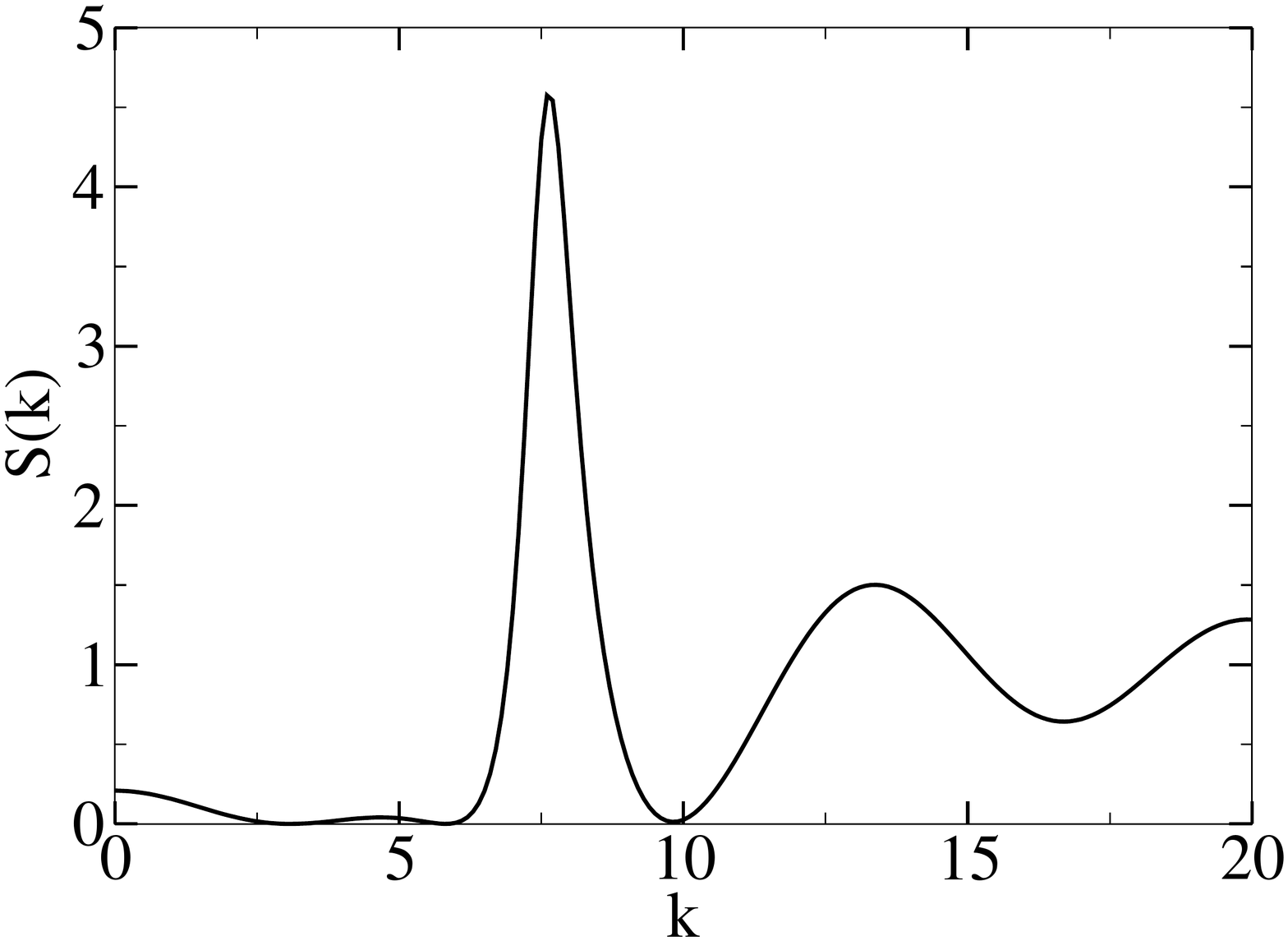}
\caption{Top: The pair correlation function $g_2(r)$ for $\phi_* = 0.64268$,
$A_1 = 3.0996$, $B_1 = 0.58091$, $C_1 = 7.54069$, $D_1 = 0.45970$, $Z = 5.0633$.
Bottom: The corresponding structure factor $S(k)$.}
\label{simpleMaxPhi1}
\end{figure}
\begin{figure}[ht]
\includegraphics[width = 4.0in,viewport = 25 30 720 
580,clip]{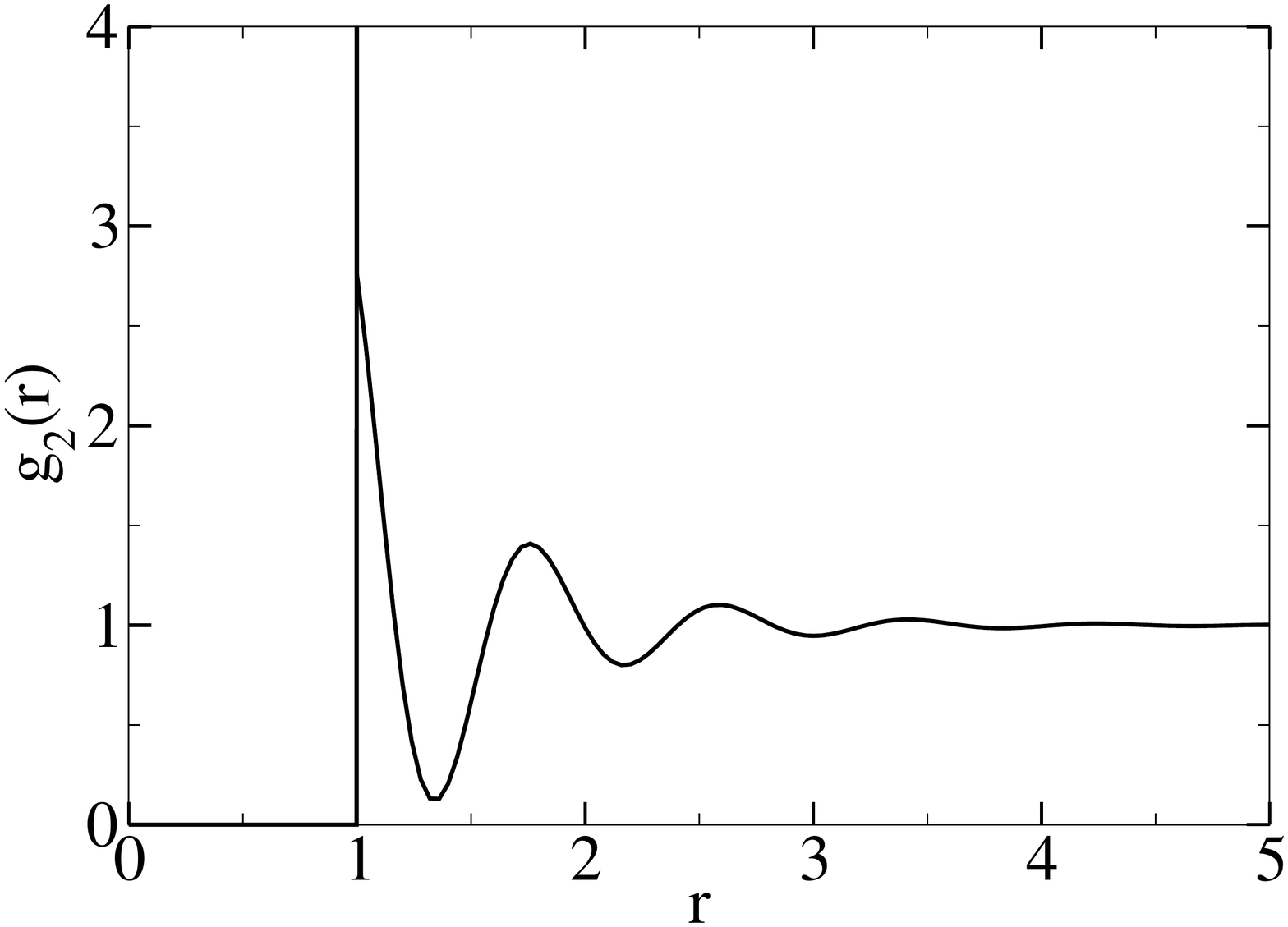}
\includegraphics[width = 4.0in,viewport = 25 30 720 
550,clip]{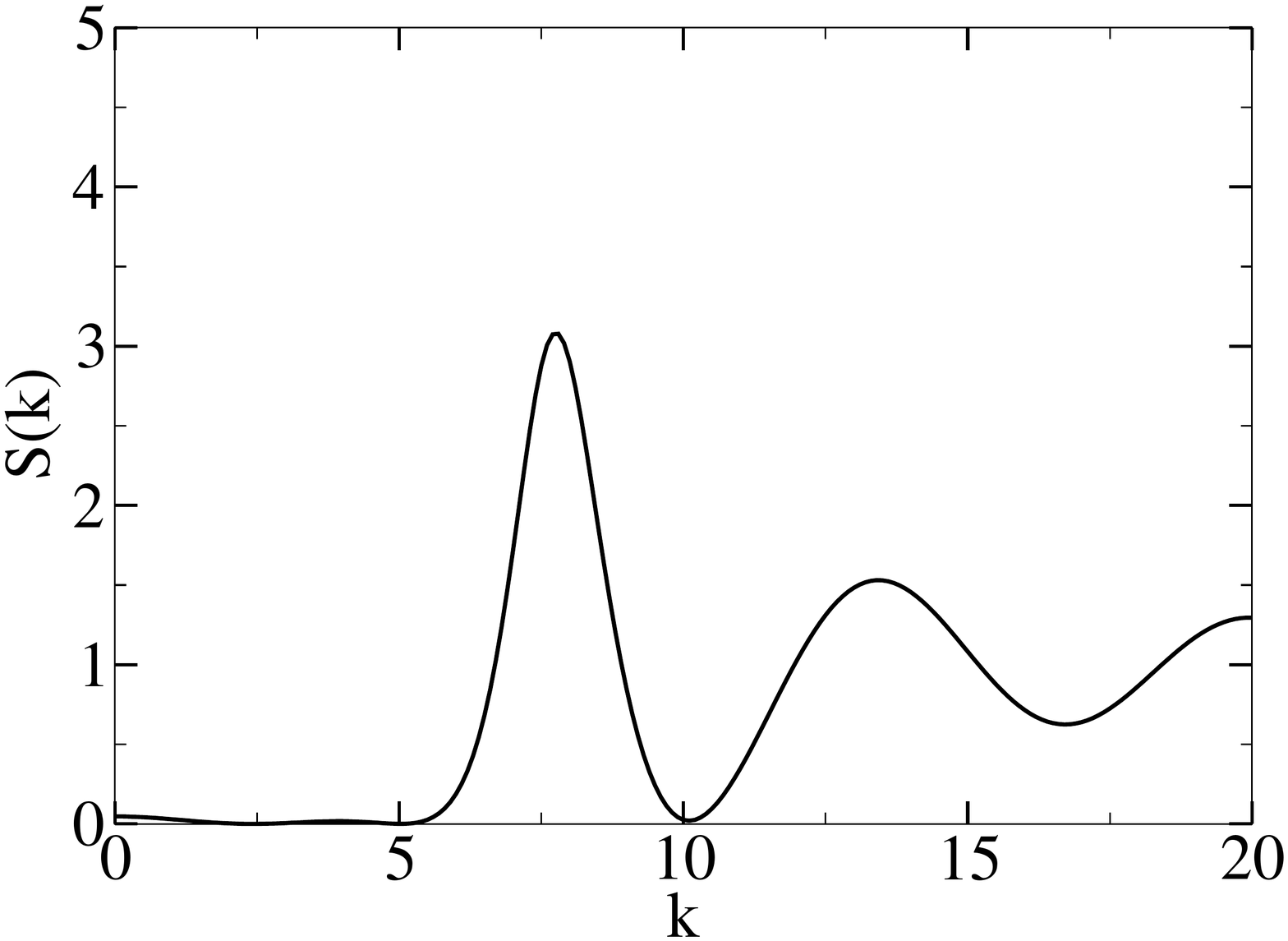}
\caption{Top: The pair correlation function $g_2(r)$ for $\phi_* = 0.64050$, 
$A_1 = 6.1707$, $B_1 = 1.2090$, $C_1 = 7.6011$, $D_1 = 0.54981$, $Z = 5.1593$.
Bottom: The corresponding structure factor $S(k)$.}
\label{simpleMaxPhi2}
\end{figure}

Figures \ref{simpleMaxPhi1} and \ref{simpleMaxPhi2} present pair
correlation functions and corresponding structure factors that yield
terminal packing fractions $\phi_* = 0.64268$ and $\phi_* = 0.64050$, respectively. 
Parameters for high packing-fraction results were similar in period $C_1$, phase 
$D_1$, and average kissing number $Z$, but varied in amplitude $A_1$ and 
damping factor $B_1$. For example, the $g_2(r)$ with the highest two terminal 
packing fractions, $\phi_* = 0.64268$ and $\phi_* = 0.64050$, exhibited 
$A_1 = 3.0996$ and $A_1 = 6.1707$, with $B_1 = 0.58091$ and $B_1 = 1.2090$, 
respectively. Each $g_2(r)$ exhibits a minimum near $r = 1.35$, with the 
minimum for the highest packing fraction equal to zero, and a maximum at 
$r = 1$ of about $2.7$, suggesting that these traits, along with a period 
$2\pi/C_1$ of about $1.2$ and a phase $D_1$ of about $0.50$ are important in 
obtaining the maximum packing fraction for this functional form. Further analysis 
indicates that the period and existence of a deep minimum within $r \approx 1.5$
remain important in maximizing packing fraction when other elements are 
added to this functional form. 

\subsection{Maximizing kissing number}

Maximum packing-fraction $g_2$'s for this functional form do not correspond 
to $g_2$'s that maximize average kissing number, as is the general case for 
sphere packings in many dimensions \cite{SST2008a}. Moreover, though the 
average kissing number for the highest possible packing fraction (FCC lattice) 
is 12 \cite{Leech1956a}, the proved maximum possible, $Z$ may 
only obtain the value of 9.5401 for this form. As will be seen later, as additional 
elements and parameters are added to this form and optimized for maximum 
packing fraction, average kissing number increases substantially. It is of 
interest therefore to maximize $Z$ with packing fraction $\phi$ as a 
parameter. The method employed is the same as before: simulated annealing in 
20,000 independent runs.

\begin{figure}[ht]
\centering
\includegraphics[width = 4.0in,viewport = 25 30 720
580,clip]{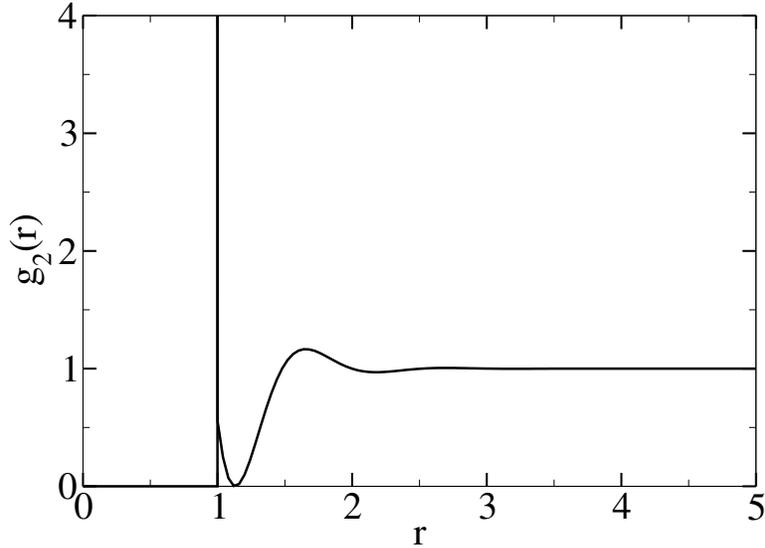}
\caption{The pair correlation function $g_2(r)$ for
$Z = 9.532$, $A_1 = -29.02$, $B_1 = 2.735$, $C_1 = 6.0537$, $D_1 = 0.47$,
$\phi = 0.631$}
\label{simpleMaxZGR}
\end{figure}

Figure \ref{simpleMaxZGR} shows the pair correlation function for the 
optimal parameters $A_1$, $B_1$, $C_1$, $D_1$, and $\phi$, with optimized $Z = 9.5401$. 
Most notably, the packing fraction associated with the maximum average kissing 
number configuration is 0.631, nearly equal to the maximum packing fraction achieved 
($\phi_* = 0.64268$) for this functional form. The fact that 
many more spheres are in contact (relative to the case
in which the packing fraction is maximized, where $Z = 5.0633$), could reflect the presence of large clusters
in the packing or even a sample-spanning cluster \cite{cluster}. Also of note 
is that the first minimum, while still equal to zero, has moved inwards 
toward $r = 1$, and that the value of the pair correlation function at 
$r = 1^+$, $g_2(1^+) = 0.5514$, is much smaller than before, where the notation 
$1^+$ refers to the right hand limit of $g_2(r)$ as $r \rightarrow 1$. 

These features must be present with an average kissing number near 12, as the 
integral of $g_2(r)$ from $r = 1^-$ to a given $R$ is related to the number 
of sphere centers most likely to be found in that range of distances, and for 
a realizable packing, this number cannot exceed 12 for $R$ near 
unity. Specifically, $Z(R)$, the total expected number of sphere centers to 
be found within a (larger) sphere of radius $R$ centered on an arbitrary 
sphere center within the packing, can be written
\begin{equation}
Z(R) = 4\pi\rho\int_{1^-}^Rx^2g_2(x)dx,
\label{zetaTot}
\end{equation}
where the average of $Z(R)$ as $R \rightarrow 1$ from the right ($1^+$) is 
equivalent to the average kissing number $Z$. 

Generally, $Z(R)$ cannot exceed some $Z_{max}(R)$, where $Z_{max}(R)$ 
represents the maximum number of sphere centers that can be placed within a 
(larger) sphere of radius $R$ centered on an arbitrary sphere center. It is 
clear from geometric considerations that $Z_{max}(R)$ is a piecewise 
continuous nowhere decreasing function of $R$. As the maximum number 
of congruent spheres that can be placed around a congruent central sphere is 
12, $Z_{max}(R)$ is $12$ on the interval $\{R: (1,1+\alpha)\}$ for some small 
deterministic parameter $\alpha > 0$, where $\alpha$ is the 
distance from unity to the first discontinuity in $Z_{max}(R)$. Currently, $\alpha$ is not known
rigorously, though its value is suspected to be about $0.045$ \cite{SvdW1951a}.
That $Z(R)$ must be less than or equal to $Z_{max}(R)$ for all $R$ is another 
necessary realizability condition for $g_2$'s representing sphere packings. 
More will be said about this condition in the conclusions.

It is also of note that the damping parameter $B_1$ for the maximum average 
kissing number case is much larger than for the maximum packing fraction 
case, implying that in the maximum average kissing number case, spatial 
correlation decreases substantially more quickly. The configuration represented by 
the more heavily damped pair correlation function can be interpreted as many 
groups of tightly packed spheres pushed together in a random fashion, thus 
exhibiting a high average kissing number but little to no spatial correlation 
at distances greater than several sphere diameters.

\section{Increasing packing fraction}

Here we show that the addition past contact of sinusoids decaying 
(to unity) as $r^{-4}$, a sinusoidally modulated form of the $r^{-4}$ decay 
present in pair correlation functions calculated from 
large-scale simulations ($10^6$ spheres) of MRJ-like packings 
\cite{DST2005a}, to the initial five-parameter family of $g_2$'s allows for 
significant increase in terminal packing fraction above 0.64268. In three 
dimensions, the inverse $4^{th}$ power is the smallest integer power to which 
a pair correlation function may decay while satisfying the $r^{-3-\epsilon}$ 
condition for representing a disordered packing. However, aside from the clues 
that this mathematical attribute might provide, the structural origins of 
the $r^{-4}$ decay currently remain a conceptual mystery.

While the form of $r^{-4}$ decay allows increased packing fraction, other 
forms, including those representative of certain other features present in 
MRJ states, do not necessarily. In fact many other additions have been 
considered for this study, though from these no substantial increases in 
$\phi_*$ above the 0.64268 achieved for the five-parameter form have been 
obtained. This is not to say that only sinusoids decaying as $r^{-4}$ will 
increase the maximum possible attainable packing fraction, just that the selection of 
elements is non-trivial. In particular, one feature present in MRJ states, 
a fractional power-law divergence near $r = 1$, resulted only in a reduced 
value for $\phi_*$. The addition of this feature is discussed in Appendix B.

The method used to optimize parameters with additional forms included 
involves three steps. Structure factors are calculated analytically from Eq. 
(\ref{FourierRadialStructFact}) as before, but to accommodate the increased 
processing time required to find the minima of $S(k)$ due to the complexity 
of its analytic form, the number of points $k$ at which $S(k)$ is calculated 
is initially reduced. The first step then is to find several ``rough" maximum 
density configurations, just as with the initial five-parameter 
configurations, using a reduced number of calculations with initial 
parameters selected as before in 10,000 independent runs. The second step is 
to improve upon these approximate maxima by increasing the number density of 
points $k$ calculated in $S(k)$ about its minimum, in 1,000 additional runs 
for each rough maximum $\phi$. The results of the second step still do not 
yield an exact maximum and hence the third step is to fine-tune the maximum 
from the second step manually, ensuring that $S(k)$ and $g_2(r)$ are indeed 
greater than zero for all $r$ and $k$.

Sinusoids decaying as $r^{-4}$ are highly successful in increasing maximum 
packing fraction beyond 0.64268. Specifically, cosine functions of the form
\begin{equation}
g_{IV}(r) = \frac{A}{r^4}\cos(Br + C)\Theta(r-1)
\label{cosQuadTail}
\end{equation}
are employed. Adding two of these elements to the initial five-parameter form 
allows identification of a function obeying all conditions with a maximum 
packing fraction of 0.6850, 45\% closer to the packing fraction of the FCC 
configuration than 0.640. Function (\ref{DCQuadTail}) is the 11-parameter 
pair correlation functional form specified.
\begin{align}
g_2(r) &= \frac{Z}{4\pi\rho}\delta(r-1) + \left(1 + \frac{A_1}{r}\exp{(-B_1r)}\sin(C_1r + D_1)\right)\Theta(r-1) \notag \\
&+ \left(\frac{A_2}{r^4}\cos(B_2r + C_2) + \frac{A_3}{r^4}\cos(B_3r + C_3)\right)\Theta(r-1), \label{DCQuadTail}
\end{align}
or
\begin{equation}
g_2(r) = g_I(r) + g_{II}(r) + g_{III}(r) + g_{IVa}(r) + g_{IVb}(r),
\label{DCQuadTail2}
\end{equation}
where the subscripts $a$ and $b$ in the last two terms on the right refer to 
the fourth and fifth terms in (\ref{DCQuadTail}). The analytical structure 
factors for the pair correlation functions represented by function 
(\ref{DCQuadTail}) calculated using relation (\ref{structFact2}) are given 
in Appendix A. 

\begin{figure}[ht]
\centering
\includegraphics[width = 4.0in,viewport = 25 30 720
580,clip]{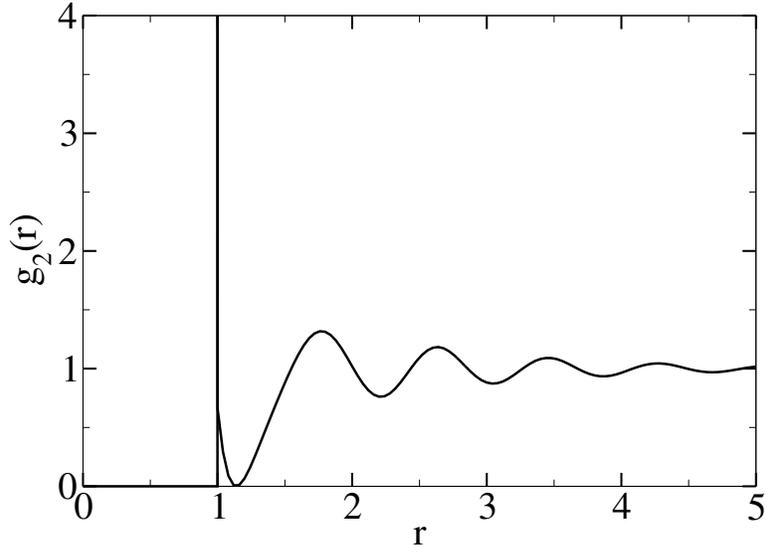}
\caption{The pair correlation function $g_2(r)$ for
$ \phi_* = 0.6850$, $A_1 = 3.555$, $B_1 = 0.7189$, $C_1 = 7.5589$, $D_1 = 0.6247$,
$Z = 10.4281$, $A_2 = 0.2205$, $B_2 = 2.370$, $C_2 = 0.000$, $A_3 = 2.2822$,
$B_3 = 8.7373$, $C_3 = 0.0423$}
\label{maxPhiGR6850}
\end{figure}
\begin{figure}[ht]
\includegraphics[width = 4.0in,viewport = 25 30 720
580,clip]{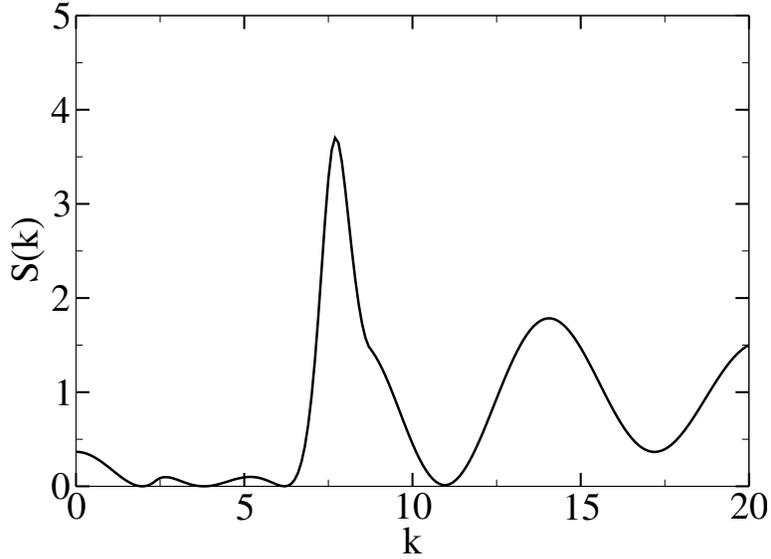}
\caption{The structure factor $S(k)$ for $ \phi_* = 0.6850$,
$A_1 = 3.555$, $B_1 = 0.7189$, $C_1 = 7.5589$, $D_1 = 0.6247$, $Z = 10.4281$,
$A_2 = 0.2205$, $B_2 = 2.370$, $C_2 = 0.000$, $A_3 = 2.2822$, $B_3 = 8.7373$,
$C_3 = 0.0423$}
\label{maxPhiSK6850}
\end{figure}

Figures (\ref{maxPhiGR6850}) and (\ref{maxPhiSK6850}) show graphs of the pair 
correlation function and structure factor with the 11 parameters of function 
(\ref{DCQuadTail}) optimized for a maximum packing fraction of 0.6850.
The first minimum and height of $g_2(r)$ just past $r = 1$ in Fig. 
\ref{maxPhiGR6850} are similar to those observed in Fig. 
\ref{simpleMaxZGR}, the plot of the pair correlation function for a 
maximized average kissing number using the initial five-parameter functional 
form. The period $2\pi/C_1$, phase $D_1$, and decay factor $B_1$ of the 
exponentially decaying sinusoidal function are similar to those observed in 
the top plots of Figs. \ref{simpleMaxPhi1} and \ref{simpleMaxPhi2}, the plots of the 
pair correlation functions for maximized packing fraction using the initial 
five-parameter functional form. The minimum and height just past $r = 1$ must 
be similar for physical meaningfulness due to high average kissing number 
$Z = 10.4281$, while the similarities in period, phase, and decay factor 
indicate that a higher degree of order is maintained farther from $r = 1$.

The decay factor, $B_1 = 0.7189$, for the 0.6850 maximum packing fraction 
function is in fact larger than the decay factor, $B_1 = 0.58091$, for the 
0.64268 packing fraction function, but fast decay in the former $g_2$ function 
is avoided through the addition of the $r^{-4}$ sinusoids.

The average kissing number $Z = 10.4281$ for the 0.6850 maximum packing 
fraction function is substantially higher than that obtained for the 0.6427 
maximum packing fraction function. The addition of the $r^{-4}$ elements 
allows not only for tighter packing of spheres (and hence a higher $Z$ than 
previously possible), but also for slower decay than present with an 
exponentially decaying function alone. This implies (but does not prove) that 
more correlation at greater distances is necessary for higher packing fraction 
disordered packings: a hypothesis that will be supported further later in this paper 
by order metric calculations.

It is of note that reducing the packing fraction $\phi$ without proportionally 
reducing the kissing number $Z$ quickly violates the structure factor 
condition. This implies that kissing numbers as high as $10.42$ cannot be 
maintained without proportionally high densities, which is in agreement with 
physical intuition: average number of spheres in contact cannot increase past 
a certain point without high enough packing fraction, though it is important to note 
that the converse of this statement is not true (if a configuration is not 
required to be jammed).

\begin{figure}[ht]
\centering
\includegraphics[width = 4.0in,viewport = 25 30 720
580,clip]{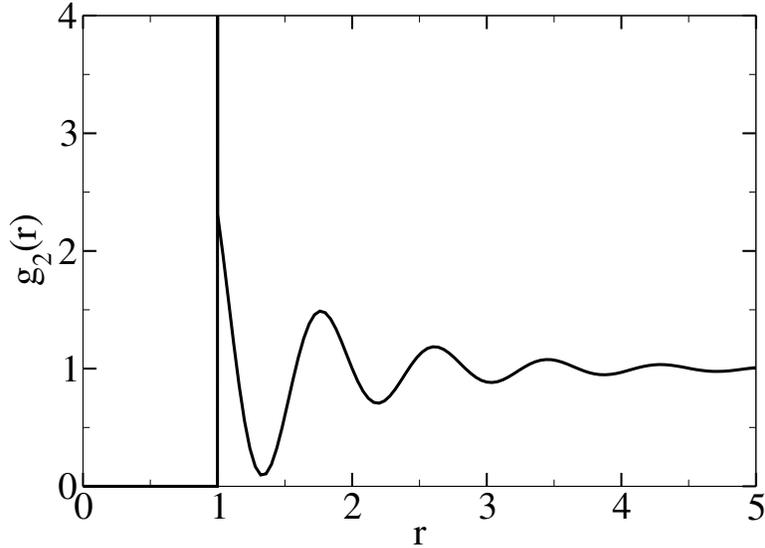}
\caption{The pair correlation function $g_2(r)$ for $ \phi_* = 0.6700$,
$A_1 = 3.3048$, $B_1 = 0.71901$, $C_1 = 7.5115$, $D_1 = 0.6610$,
$Z = 6.672$, $A_2 = 0.0795$, $B_2 = 2.390$, $C_2 = 0.000$, $A_3 = 0.312$,
$B_3 = 8.389$, $C_3 = 0.000$}
\label{maxPhiGR6700}
\end{figure}

Further supporting this notion is Fig. \ref{maxPhiGR6700}, a graph of the 
pair correlation function with 10 parameters optimized to maximize $\phi$ 
with $Z$ set to $6.672$.
With the four parameters of function (\ref{g3}) nearly the same as for the 
0.6850 terminal packing fraction function but $Z$ reduced substantially, only 
a lower terminal packing fraction is possible. The amplitudes of the $r^{-4}$ 
cosine elements are also substantially reduced in Fig. \ref{maxPhiGR6700}, 
implying that the increase in packing fraction from 0.6700 to 0.6850 necessitates both 
correlation at greater distances and a substantially higher average kissing 
number. As will be quantified later in this paper, it follows that more order 
overall is required to attain the packing fraction 0.6850; nonetheless, the 
smooth correlation function past contact and swift decay to unity indicates 
that the configuration remains disordered as defined.

As before with the five-parameter functional form, it is of interest
to maximize kissing number $Z$ for the 11-parameter functional form
(\ref{DCQuadTail}).
We find that the average kissing number cannot increase much above 10.4281. For this functional form
and with realizability conditions satisfied, we see
that  if one of either the packing fraction or kissing number is near its maximum value, then the other must also be near its maximum.

\section{From disordered to crystalline}
\subsection{Order metrics}
A disordered packing as defined clearly lacks the degree of order present in 
a periodic packing, but the degree of order within a disordered system still 
varies, and may be quantified. For cases of hard-sphere packings, many 
different methods of quantifying the order in a configuration are possible. 
The majority of these methods advocate the use of scalar statistical 
measures. Some of these assume that the most ordered system
is the appropriate maximally dense  packing \cite{TTD2000a,SNR1983a,TTD2000b,Ka02,TorquatoRHM2002},
while others do not presuppose a reference crystal state \cite{TTD2000b,Ka02,TorquatoRHM2002,TS2003a,TDS2006a}.

For our purposes, the translational order metric introduced
by Truskett \textit{et al.} \cite{TTD2000b} is convenient because
it is given in terms of the total correlation function:
\begin{equation}
\mbox{$\cal{T}$} \equiv \frac{1}{\chi_c - \rho^{1/3}D}\int_{\rho^{1/3}D}^{\chi_c}|h(\chi\rho^{-1/3})|d\chi,
\label{TransOrderParam}
\end{equation}
where $D = 1$ is the diameter of the hard spheres, $\chi = r\rho^{1/3}$, and 
$\chi_c$ is a selected cutoff distance \cite{TTD2000b}. The rescaled radial 
coordinate $\chi$ is set such that packing fractions of varying densities 
will be comparable, in that the total number of sphere centers over the 
integration range in packings of varying densities will be the same.
The quantity $\chi_c$ must be chosen with care: in a short-ranged disordered 
system, if $\chi_c$ is chosen too large, the metric will not discriminate 
well between states as $h(\chi) \rightarrow 0$ with $\chi \rightarrow \infty$.
If $\chi_c$ is chosen too small, $\cal{T}$ will not take contributions 
from correlations at greater $r$ into account. For the purpose 
of measuring order in all of the relatively short-range pair correlation 
functions described in this paper, $\chi_c = 4$ (or for packing fraction near 
0.65, about $r = 3.75$) is chosen to ensure that all fluctuations of the 
total correlation functions about zero greater than a certain minimal 
amplitude (about 0.1) are taken into account.


Using the translational order metric (\ref{TransOrderParam}), we show here that the 
degree of order in the 
configurations presented is greatest for the maximum packing fraction achieved. Table 
\ref{orderTable}, referenced by figure number and $g_2(r)$ functional form, 
provies the average kissing number $Z$ and terminal packing fraction $\phi_*$ 
versus order $\cal{T}$ for the pair correlation functions previously 
discussed. The table conveys that large differences in order may exist for similar 
densities, as is the case with the pair correlation functions represented by 
the top plots of Figs. \ref{simpleMaxPhi1} and \ref{simpleMaxPhi2}. Additionally, a
higher average kissing number does not necessitate higher order: the degree 
of order in the maximum average kissing number case of the five parameter 
form (Fig. \ref{simpleMaxZGR}) is substantially lower than in the the 
maximum packing fraction case (Fig. \ref{simpleMaxPhi1}). Here, greater 
correlation at a distance is sacrificed to put more particles in contact. 
Finally, it is of note that for the maximum packing fraction cases of the 
five- and 11-parameter forms, represented by the top plot of Fig. \ref{simpleMaxPhi1}, and 
Fig. \ref{maxPhiGR6850}, respectively, order is greatest. These results are by no 
means conclusive, but do suggest that the maximum packing fraction and maximum order 
representations of a given pair correlation function functional form are 
similar or even perhaps the same. 

\begin{table}[ht]
\caption{$\cal{T}$ order metric values}
\centering
\begin{tabular}{c c c c c}
\hline\hline
\qquad $\phi_*$ \qquad & \qquad $Z$ \qquad & \qquad $\cal{T}$ \qquad & \qquad 
Fig. \,\# \qquad & \qquad Eq. \,\# \qquad \\
\hline
\qquad 0.631 \qquad & \qquad 9.532 \qquad & \qquad 0.36 \qquad & \qquad 
\ref{simpleMaxZGR} \qquad & \qquad \ref{fiveForm} \qquad \\
\qquad 0.64050 \qquad & \qquad 5.1593 \qquad & \qquad 0.34 \qquad & \qquad 
\ref{simpleMaxPhi2} \qquad & \qquad \ref{fiveForm} \qquad \\
\qquad 0.64268 \qquad & \qquad 5.0633 \qquad & \qquad 0.43 \qquad & \qquad 
\ref{simpleMaxPhi1} \qquad & \qquad \ref{fiveForm} \qquad \\
\qquad 0.670 \qquad & \qquad 6.672 \qquad & \qquad 0.40 \qquad & \qquad 
\ref{maxPhiGR6700} \qquad & \qquad \ref{DCQuadTail2} \qquad \\
\qquad 0.6850 \qquad & \qquad 10.4281 \qquad & \qquad 0.46 \qquad & \qquad 
\ref{maxPhiGR6850} \qquad & \qquad \ref{DCQuadTail2} \qquad \\
\hline
\end{tabular}
\label{orderTable}
\end{table}

\subsection{Moving toward crystalline order}

In the optimization study above only two sinusoids decaying as $r^{-4}$ were 
added to the initial five parameter form. We now show that an infinite (or 
finite) number of sinusoids decaying as $r^{-4}$, along with a Heaviside step
function centered at $r=1$, can represent any bounded (nowhere infinite) 
piecewise-differentiable pair correlation function that decays to unity sufficiently fast. This 
follows directly from the existence of Fourier transforms, as will be explained 
shortly in greater detail, assuming that we explicitly state the bounding and 
decay conditions on the pair correlation function to mean that the Fourier transform
of $r^4f(r) = r^4(g_2(r) - \Theta(r-1))$ exists.



From a Fourier integral theorem, a radial function \cite{endnote4} 
$f(r)$ can be represented on an infinite interval not containing zero, for our purposes, 
$[1,\infty)$, by a continuum of sinusoids decaying as $r^{-4}$, so long as the
one-dimensional Fourier transform of $r^4f(r)$ over the interval exists in the 
sense of generalized functions \cite{Churchill1956}, as follows:
\begin{equation}
f(r) = 
\begin{cases}
\int_{0}^{\infty}\tilde{F}(k)\frac{\cos{(kr)}}{r^4}dk, & r\geq 1 \\
0, & r < 1 \\
\end{cases}
\label{FourierR4}
\end{equation}
with
\begin{equation}
\tilde{F}(k) = \frac{1}{\pi}\int_1^{\infty}x^4f(x)\cos{(kx)}dx.
\label{ckr}
\end{equation}
Under these conditions, a smooth disordered hard-sphere 
$g_2$ that decays to unity can be written simply as
\begin{equation}
g_2(r) = f(r) + \Theta(r-1).
\label{g2Fourier}
\end{equation}
For example, to represent $g_{IV}(r)$ (Eq. (\ref{cosQuadTail})) in this form with $C = 0$, 
using the parameters from Eq. (\ref{cosQuadTail}), we set $\tilde{F}(k) = A\delta(k-B)$.

Discretizing the $k$ in expression (\ref{ckr}) with increment $\Delta k$ between 
two successive discrete values, $f(r)$ may be approximated with $\tilde{F}(k)$ a series of real constants. In the limit as 
$\Delta k \rightarrow 0$ or as $k$ becomes continuous, expression (\ref{FourierR4}) holds and any 
$g_2(r)$ meeting the aforementioned conditions may be represented by Eq. (\ref{g2Fourier}).
Thus, to demonstrate that disordered packings as defined can attain densities up to $\pi/\sqrt{18}$, 
it only remains to show that there is a realizable packing with packing fraction near $\pi/\sqrt{18}$ 
that can be represented by a pair correlation function in the form of Eq. (\ref{g2Fourier}) for 
which the Fourier transform of $r^4f(r)$ exists, which is the focus of the remainder of this section.

The pair correlation function of an infinite close-packed FCC crystal 
consisting of congruent spheres of diameter $D = 1$ can be written as
\begin{equation}
g_2^{fcc}(r) = \frac{1}{4\pi\rho}\sum_{i = 1}^{\infty}\frac{C_i^{fcc}}{q_i^2}\delta(r - q_i),
\label{FCCg2}
\end{equation}
where the $q_i$ are the distances from the origin to each successive shell in an 
FCC packing (which in these scaled coordinates are $q_i = \sqrt{i}$), the sum runs 
only over those $i$ for which shells are present (for example, $i = 14$, $30$, $46$, 
etc. are skipped), the $C_i^{fcc}$ are the coefficients of the (even-exponent) 
terms in the FCC theta series \cite{CSSPLG1999} (which represent the number 
of spheres present on a shell), and $\delta$ represents a Dirac delta 
function. Any same-density stacking variant of an FCC configuration, 
i.e., any of the Barlow packings \cite{Barlow1883a,endnote3}, may be represented by 
a similar series of delta functions. For a spherical crystal of finite size, 
the $q_i$ remain the same but terminate for $i > 2I$ where $\sqrt{I}$ is the 
distance from the central sphere center to the center of the outermost sphere 
of the crystal, and the $C_i$ must be reduced to reflect a lower average 
number of spheres present at given distances for all but the central sphere.

An FCC configuration is not, however, disordered. To create a disordered packing
from a series of finite-size FCC crystals, three further steps must be
taken. First, the delta functions are replaced with smooth, sharply peaked
Gaussian-like curves that decay to zero at distance $|w/2|$ from each $q_i$.
We define the integral under each of these curves from 
$q_i - w/2$ to $q_i + w/2$
to be equal to the reduced $C_i$ of a FCC crystal of specified finite size.
Physically, this represents packing spheres of diameter 1 in a FCC arrangement,
shrinking the radius of these spheres to $r_a = (1/2) - (1/4)w$, and then moving
each sphere a distance of no more than $(1/4)w$ from its initial position such
that the pair correlation function of the packing near each $q_i$ has the
properties just described. It is of note that the exact form of the pair
correlation function is not of consequence as long as the proscribed properties
are maintained. 

Second, the $C_i$ are again altered such that the finite-size
``shrunk" FCC configuration of spheres becomes cubical in shape, with
a cube of side length $I^{cube}$ circumscribing the arrangement. Finally, an infinite
number of these cubes, with principal axes of the FCC-packed
spheres within each cube arranged at uncorrelated angles, are packed together 
tightly such that each pair of parallel
faces of every cube are parallel to a pair of faces on every 
other cube, but with each face touching four other faces (i.e., four 
other cubes). Cubes may be tightly packed in this fashion such that they 
fill space, and such that the resulting pair correlation function of the spheres 
comprising these cubes is disordered due to the packing of the cubes and the 
random arrangement of the principal axes of the FCC-packed spheres within each cube.

This pair correlation function would be in practice difficult to write 
explicitly due to boundary effects, but this is inconsequential to the 
description, as all that is necessary is to illustrate that such packings exist.
Additionally, it is noteworthy that initially packing the spheres in each cube in
arrangements equivalent to any of the Barlow packings would yield an equivalent result.

Boundary effects become negligible as the number of cubes approaches 
infinity, and for increasing $I^{cube}$ and decreasing $w$, any packing 
fraction $\pi/6 < \phi < \pi/\sqrt{18}$, with $\pi/6 = \phi^{sc}$ the packing 
fraction of a simple cubic arrangement, can be obtained. The minimal packing fraction
$\phi^{sc}$ is obtained when only one sphere is present in each cube. 
The distance  $w$ can be made small enough for any finite-sized system of cubes
such that the system exhibits physical stability when force is applied,
though the pair correlation function of such a system would 
change (perhaps to include delta functions) as soon as stress and strain 
were present.

In this way, as long as $w > 0$ and $I^{cube}$ is finite, a disordered 
packing may exhibit any packing fraction up to $\pi/\sqrt{18}$, if one accepts
that the pair correlation function of the system described above, represented
in the form of Eq. (\ref{g2Fourier}), decays to unity at least as fast as 
$r^{-3-\epsilon}$ and in a form such that the Fourier
transform of $r^4f(r)$ exists. Due to this result and the others
demonstrated in this paper, we state that the sequential addition 
and optimization of more than two sinusoidal terms of the form of expression 
(\ref{cosQuadTail}) will allow packing fractions greater than 0.6850, and as 
the number of terms grows, the optimal packing fraction for a realizable 
$g_2$ is conjectured to approach $\pi/\sqrt{18}$.

\section{Conclusions and Discussion}

Using the $g_2$-invariant method with $g_2$'s satisfying the three necessary, 
but generally not sufficient, conditions for realizability, we demonstrated 
without implicit reliance upon any packing methodology that packing fractions 
well above $0.64$ are obtainable for pair correlation functions incorporating 
the salient features of disordered packings. A packing fraction of 0.6850 was 
obtained employing a test family of $g_2$'s mimicking features observed in MRJ 
packings, including most notably core exclusion, contact pairs, and a 
sinusoidal decay to unity as $r^{-4}$. Consistent with a previous study 
\cite{TTD2000a}, we found that to achieve higher packing fractions, the 
degree of order must increase.

Additionally we showed, employing a qualitative example and a revised
version of a Fourier integral theorem, the surprising result that a disordered 
packing as defined may reach packing fractions approaching $\pi/\sqrt{18}$, the 
maximum possible for a three-dimensional hard-sphere packing. These results 
support the hypothesis that continued addition and subsequent optimization of 
sinusoids decaying as $r^{-4}$ (of the form of Eq. (\ref{cosQuadTail})) will find 
realizable $g_2$ with higher terminal packing fractions up to (but not reaching) 
$\pi/\sqrt{18}$.

The conclusion that the addition of sinusoids decaying as $r^{-4}$ allows 
higher terminal packing fractions (where the addition of other features does 
not necessarily) is very relevant to the study of high-density physical and 
jammed disordered systems, since it demonstrates that this 
feature contributes significantly in allowing the systems to reach 
aforementioned higher densities. It is noteworthy that the $r^{-4}$ decay to unity present in 
the $g_2$'s of MRJ packings also characterizes $g_2$'s of high-density Bose 
systems \cite{RC1967a}, ground states of fermionic systems \cite{To08},
and models of the density distribution of the early 
Universe \cite{Harrison1970a,Zeldovich1972a,PY1970a,GJL2002a}, though these
latter systems are not sphere packings.

In future work, we will seek to extend these results to configurations of 
multi-component spheres and non-spherical objects. Additionally, we will 
investigate the structural origins of the presence of the $r^{-4}$ decay in 
MRJ states, and attempt to determine if in higher dimensions additions of 
terms consisting of sinusoids decaying as the smallest (inverse) integer 
power for which a packing remains disordered, $r^{-d-1}$, to a $g_2(r)$ 
representing the salient features of a $d$-dimensional hard sphere system, will 
allow for packing fractions up to the known maximum in that dimension.

Moreover, we will examine the form of $Z_{max}(R)$, the maximum number of 
sphere centers that can be placed within a (larger) sphere of radius $R$ 
centered on an arbitrary sphere center. A natural question to ask is whether 
the realizability condition $Z(R) \leq Z_{max}(R)$ further constrains 
$g_2(r)$ beyond the three conditions (\ref{condition1}), (\ref{condition2}), 
and (\ref{condition3}). The answer is in the affirmative. We have already 
noted that the Cohn-Elkies linear programming upper bound formulation 
\cite{CE2003a} is the dual of the Torquato-Stillinger lower bound procedure, 
i.e., the $g_2$-invariant process \cite{TS2006a}. Cohn and Kumar 
\cite{CK2009a} recently proved that there is no duality gap between the 
primal and dual LP programs, which means both the upper and lower bounds 
coincide when the best test functions are employed. Cohn and Elkies were able 
to find the test functions that yield the best upper bound on the maximal 
packing fraction in three dimensions: a packing fraction of about 0.778, which 
is well above the true maximal value. This means that there exists a test 
pair correlation function for the lower bound formulation that will deliver 
the same maximal packing fraction of 0.778, which clearly is not realizable. 
It was shown elsewhere that applying the additional condition that $Z_{max}$ 
must be equal to 12 up to some small positive $\alpha$ beyond contact together 
with the best test function in the upper bound brought down the maximal 
packing fraction appreciably from 0.778 \cite{CKT2009a}. This means that 
adding the $Z_{max}$ condition to the corresponding best test pair 
correlation function will also improve the packing fraction estimate. Therefore, the 
$Z_{max}$ condition introduces additional information beyond that contained 
in the two standard nonnegativity conditions. The precise form of this
additional condition has yet to be fully elucidated and will  be explored in future work.

\begin{acknowledgments}
The authors are grateful to Henry Cohn for a critical reading of our manuscript.
S.T. thanks the Institute for Advanced Study for its hospitality during his 
stay there. This work was supported by the Division of Mathematical Sciences 
at the National Science Foundation under Award Number DMS-0804431
and by the MRSEC Program of the National Science Foundation under Award Number DMR-0820341.
\end{acknowledgments}

\appendix

\section{Analytical structure factor components}

The analytical structure factor components, as calculated from relation 
(\ref{structFact2}), for $g_I$, $g_{II}$, $g_{III}$, $g_{IV}$ and $g_V$, 
with $r = 1$ the diameter of the spheres.

\small
\begin{equation}
\mathbf{G_I(}k\mathbf{)} = \frac{4\pi}{k^3}(k\cos{k}-\sin{k})
\label{GI}
\end{equation}
\begin{equation}
\mathbf{G_{II}(}k\mathbf{)} = \frac{Z}{\rho k}\sin{k}
\label{GII}
\end{equation}
\begin{align}
&\mathbf{G_{III}(}k\mathbf{)} = \frac{2\pi Ae^{-B}}{k} \notag \\
&\bigg(\frac{B\cos{(k-C-D)} - (k-C)\sin{(k-C-D)}}{B^2 + (k-C)^2} - \notag \\
&\frac{B\cos{(k+C+D)} - (k+C)\sin{(k+C+D)}}{B^2 + (k+C)^2}\bigg) \label{GIII} \\
\notag
\end{align}
\begin{align}
&\mathbf{G_{IV}(}k\mathbf{)} = \frac{A\pi}{2k}\Big(\cos{C}\big[\pi(B-k)|B-k| - \pi(B+k)^2 \notag \\
&+ 2(B+k)^2\mbox{Si}(B+k) - 2(B-k)^2\mbox{Si}(B-k)\big] \notag \\
&+ \sin{C}\big[2(B+k)^2\mbox{Ci}(B+k) + (B-k)^2\big(2\log{(B-k)} \notag \\
&- \log{(B-k)^2} - 2\mbox{Ci}(B-k)\big)\big] + 4k\cos{k}\cos{(B+C)} \notag \\
&+ 4\sin{k}\big[\cos{(B+C)} - B\sin{(B+C)}\big] \Big) \label{GIV} \\
\notag
\end{align}
\begin{align}
&\mathbf{G_V(}k\mathbf{)} = 
\frac{8A\sqrt{B}\pi}{15}\Big(kB\cos{k}\big[5\,_1\mbox{F}_2\big(\frac{3}{4};\frac{3}{2},
\frac{7}{4};-\frac{1}{4}k^2B^2\big) \notag \\
&+ 3B\,_1\mbox{F}_2\big(\frac{5}{4};\frac{3}{2},\frac{9}{4};-
\frac{1}{4}k^2B^2\big)\big] + 
\big[3\,_1\mbox{F}_2\big(\frac{1}{4};\frac{1}{2},\frac{5}{4};-
\frac{1}{4}k^2B^2\big) \notag \\
&+ B\,_1\mbox{F}_2\big(\frac{3}{4};\frac{1}{2},\frac{7}{4};-
\frac{1}{4}k^2B^2\big)\big]5\sin{k}\Big) \label{GV} \\
\notag
\end{align}

\normalsize
In the expression for $G_{IV}(k)$, Si$(k)$ represents the standard sine 
integral, Si$(k) = \int_0^kdx\sin{x}/x$, and Ci$(k)$ represents the 
standard cosine integral, Ci$(k) = -\int_k^{\infty}dx\cos{x}/x$. In the 
expression for $G_V(k)$, $_p\mbox{F}_q\big(\{a_i\},\{b_j\},k\big)$ 
represents the standard hypergeometric $_p\mbox{F}_q$ function, 
\begin{equation*}
_p\mbox{F}_q\big(\{a_i\},\{b_j\},k\big) = 
\sum_{n=0}^{\infty}\frac{(a_1)_n\cdots(a_p)_nk^n}{(b_1)_n\cdots(b_q)_nn!},
\end{equation*}
with $(a_i) = \gamma(a + i)/\gamma(a)$ and $\gamma(x)$ the standard gamma 
function.

\section{Power-Law Divergence in Near Contact Distribution}

\begin{figure}[ht]
\centering
\includegraphics[width = 4.0in,viewport = 5 16 720 
580,clip]{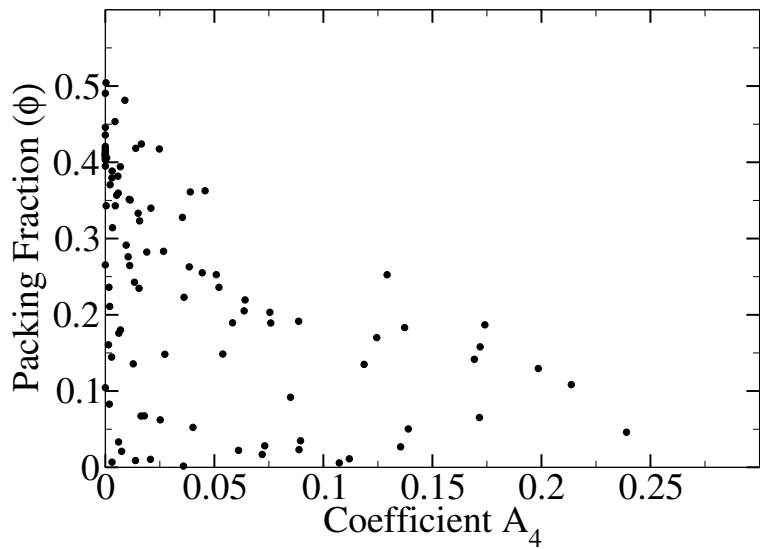}
\caption{Maximum obtained packing fraction $\phi$ versus value 
of parameter $A_4$, represented by a random selection of 100 from 10,000 
maximum density parameter optimization runs where a term $g_V(r)$, given by relation
(\ref{g5}), was added to the initial five-parameter pair correlation function 
functional form \ref{fiveForm}}
\label{inverseOneHalf}
\end{figure}

Evidence indicates that the addition of one of the salient features observed 
in MRJ-like states for three-dimensional hard-sphere packings, a
fractional power-law  divergence near $r = 1$ due to near contacts \cite{Si02,Don05}, 
does not increase packing fraction past that 
obtained from the five-parameter form. For example, the addition of the form
\begin{equation}
g_{V}(r) = \frac{A_4}{(r-1)^{1/2}}\Theta(r-1)\Theta(B_4-r),
\label{g5}
\end{equation}
with $B_4 = 1.15$ a cutoff parameter to the square-root power decay, to the 
$g_I(r)$, $g_{II}(r)$, $g_{III}(r)$ discussed earlier led to no additional 
increase in packing fraction under maximum density parameter optimization in 
10,000 independent runs. Figure \ref{inverseOneHalf} shows a graph of the maximum 
packing fraction obtained versus the value of coefficient $A_4$, where values plotted 
represent a random sample of 100 from the 10,000 runs conducted. One can see 
clearly that as $A_4$ increases in value, indicative of the increased 
prominence of the inverse square-root divergence near $r=1$, maximum packing fraction 
obtained decreases steadily.

These results imply that the inverse half power decay near $r = 1$ is a
feature intrinsic to MRJ configurations and that this feature must be diminished 
to increase packing fraction. Physically this is intuitive, as 
the presence of a half power decay near $r = 1$ indicates that there are many 
spheres smoothly distributed just outside of contact, i.e., that locally on 
average there is room to compress the system further.


\begin{thebibliography}{37}
\bibitem[{end({\natexlab{a}})}]{endnote2}
\bibinfo{note}{The requirement for smoothness past contact can be expressed
  mathematically by stating that $g_2(r)$ is a $C^n$ function with $n \geq 
1$,
  or that at least its first derivative exists for $r > 1$}.
\bibitem{cluster}
A cluster is a group of contacting particles in which there is a connected
path lying inside all of the particles of the group that never
enters the space exterior to the particles.
\bibitem{endnote3}
The Barlow packings include all permutations of the stackings of three specific planes of
spheres, $A$, $B$, $C$, where none of the three stackings is adjacent to its like (no $AA$, 
$BB$, $CC$), of which FCC and HCP are special cases, stacked $ABCABC\dots$ and $ABAB\dots$,
respectively.
\end{thebibliography}
\end{document}